\documentclass{article}

\usepackage{arxiv}

\usepackage[utf8]{inputenc} % allow utf-8 input
\usepackage[T1]{fontenc}    % use 8-bit T1 fonts
\usepackage{hyperref}       % hyperlinks
\usepackage{url}            % simple URL typesetting
\usepackage{booktabs}       % professional-quality tables
\usepackage{amsfonts}       % blackboard math symbols
\usepackage{nicefrac}       % compact symbols for 1/2, etc.
\usepackage{microtype}      % microtypography
\usepackage{lipsum}		% Can be removed after putting your text content
\usepackage{graphicx}
\usepackage{doi}

\usepackage{amsmath}
\usepackage{amssymb}
\usepackage{subcaption}
\usepackage[verbose]{placeins}
\usepackage{graphicx}% Include figure files
\usepackage{dcolumn}% Align table columns on decimal point
\usepackage{bm}% bold math
\usepackage[utf8]{inputenc}
\usepackage[T1]{fontenc}
\usepackage{mathptmx}

\title{Correlated Power Time Series of Individual Wind Turbines: a Data Driven Model Approach}

\author{ \href{https://orcid.org/0000-0002-3095-8960}{\includegraphics[scale=0.06]{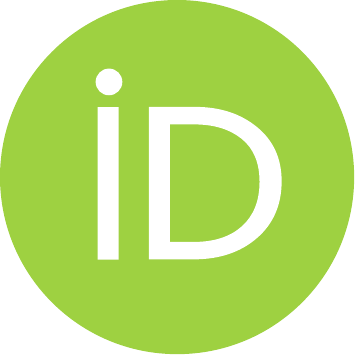}\hspace{1mm}Tobias Braun}
\thanks{Complexity Science (RD4),
	Potsdam Institute for Climate Impact Research,
	14473 Potsdam, Germany,
	Tel.: +49-331-28820744,
	\texttt{tobraun@pik-potsdam.de}}\\
	Faculty of Physics\\
	University of Duisburg-Essen\\
	Lotharstrasse 1, 47057 Duisburg, Germany\\
	\And
{\hspace{1mm}Matthias Waechter} \\
	ForWind $\&$ Institute of Physics\\
	Carl-von-Ossietzky University Oldenburg \\
	Küpkersweg 70, 26129 Oldenburg, Germany
	\And
{\hspace{1mm}Joachim Peinke} \\
	ForWind $\&$ Institute of Physics\\
	Carl-von-Ossietzky University Oldenburg \\
	Küpkersweg 70, 26129 Oldenburg, Germany
	\And
{\hspace{1mm}Thomas Guhr} \\
	Faculty of Physics \\
	University of Duisburg-Essen \\
	Lotharstrasse 1, 47057 Duisburg, Germany
}
\renewcommand{\headeright}{ }
\renewcommand{\undertitle}{ }
\renewcommand{\shorttitle}{Correlated Power Time Series of Individual Wind Turbines: a Data Driven Model Approach}

\hypersetup{
pdftitle={Correlated Power Time Series of Individual Wind Turbines: a Data Driven Model Approach},
pdfsubject={q-nlin.CD},
pdfauthor={Tobias Braun, Matthias Waechter, Joachim Peinke, Thomas Guhr},
pdfkeywords={wind energy, power generation, stochastic modelling, correlation, detrended fluctuation analysis},
}

\begin{document}
\maketitle

\begin{abstract}
Wind farms can be regarded as complex systems that are on the one hand coupled
to the nonlinear, stochastic characteristics of weather and on the other hand strongly
influenced by supervisory control mechanisms. One crucial problem in this context today is
the predictability of wind energy as an intermittent renewable resource with additional non--stationary nature. In this context, we analyse power time series measured in an offshore wind farm for a total period of one year with a time resolution of ten minutes. Applying Detrended Fluctuation Analysis, we characterize the autocorrelation of power time series and find a Hurst exponent in the persistent regime with crossover behaviour.
To enrich the modelling perspective of complex large wind energy systems, we develop a stochastic reduced--form model of power time series. Observed transitions between two dominating power generation phases are reflected by a bistable deterministic component while correlated stochastic fluctuations account for the identified persistence. The model succeeds to qualitatively reproduce several empirical characteristics such as the autocorrelation function and the bimodal probability density function.
\end{abstract}

% keywords can be removed

\section{Introduction}
In the context of anthropogenic climate change, the challenge of reducing carbon emissions is of central importance. Renewable sources of energy are considered to be one of the most promising solutions in the electricity sector to cover an increasing energy demand without exacerbating the high carbon emissions coupled to it. Wind energy in particular appears to be one of the most strongly increasing sources of renewable energy \cite{globalreport, windpotential} but demands an extraordinary adaption of grids and related power systems due to its intermittent nature \cite{crossley2010smart,carrasco2006power}. It consequently raises the need for a profound understanding of this intermittency and the opportunity to perform extensive studies on the reliability of power systems by suitable models. Such models can only be calibrated with respect to empirical data while different approaches are required to reflect features on multiple spatial and temporal scales  \cite{jiang2017review, soman2010review}. Moreover, they can be used to study the impact of certain well known statistical characteristics on the resulting dynamics. This provides wind farm controllers with a rich set of tools to analyse variable scenarios taking influences into account that are known to be of importance.

In this work, we focus on the dynamics of single wind turbine (WT) power generation in an offshore wind farm. While especially power generation of aggregated wind farms or even complexes of several wind farms have received considerable attention, the challenge of intermittent and stochastic characteristics is particularly high on the scale of single turbines \cite{ren2018analysis}. Still, it does not vanish for larger units like wind farms or national wide wind power generation \cite{anvari2016short}. How generated power fluctuations from geographically separated wind farms are dampened when aggregated \cite{martin2015variability, katzenstein2010variability} is of crucial importance for large scale grid stability \cite{heitzig2014interdisciplinary, gorjao2019data}.

Our perspective on this problem is to study the autocorrelation of power time series as an indicator of their intermittent and nonstationary nature in a first step. In most modelling approaches, information on the complex temporal evolution of variations is included and hence autocorrelation is of great general importance. Nonstationarity is another feature that is frequently encountered dealing with complex systems \cite{yuriy}. We address this within the analysis of power output correlations using the popular method of Detrended Fluctuation Analysis (DFA) \cite{kantelhardtdfa}. It can deal with polynomial nonstationarities in a generalized fashion. The fluctuations around these polynomial trends are subject to the correlation analysis which yields a Hurst exponent as an indicator for the strength and nature of the autocorrelation.

The idea that power time series should exhibit scaling behaviour is based on the finding that the underlying wind speed dynamics are governed by atmospherical turbulence \cite{kolmogorov1941local}. The traditional tool to capture this aspect is the power spectrum obtained by Fourier analysis \cite{frisch1995turbulence}. Several studies have examined the power spectral density both for wind speed \cite{calif2012pdf, calif2012modeling} and power time series \cite{baile2010intermittency, milan}, also uncovering how aggregation on different spatial scales impacts power fluctuations \cite{fertig2012effect, katzenstein2010variability}.
Other classical methods range from structure functions to estimators for the Hurst exponent based on scale dependent variance calculation on the respective time series \cite{taqqu1995estimators}. Still, these methods do not take the above mentioned nonstationarity into account.
To this extent, several works study the autocorrelation of wind speed time series $u(t)$ applying DFA and Multifractal DFA (MFDFA) and conclude that it behaves persistent with a multifractal dependence \cite{multifractalwindspeed, lrdmfwindspeed, mfspec}.
While most research in this context focuses on local wind speed measurements of single sites, some studies reveal multifractality of wind speed on a larger spatial scale. For instance, \cite{wei2018multidimensional} incorporates wind speed data spatially distributed all over Switzerland and finds multifractal dependence of persistence for the cooperative behaviour in the context of monitoring systems. Similar research for power time series appears more limited. The authors of \cite{calif2014multiscaling, turbulent} reveal a high degree of multifractality for both wind speed and power time series of an aggregated wind farm and join both in a description via generalized correlation functions. 
Furthermore, aggregated power output of wind farms is known to show complex correlations in terms of persistence and multifractality. The authors in \cite{multifractalpower} uncover multifractality for power time series of an aggregated wind farm in South Australia with data on a similar time scale. They classify power time series in the persistent regime, especially on short time scales of several minutes. Yet to our best knowledge, no research has employed DFA or multifractal methods focusing on single WTs' power output.

After we have obtained a better understanding of autocorrelations, we put forward a model based on these theoretical insights and an important feature of the empirical probability density function (PDF) of power output. A broad range of approaches is considered for wind power generation models in the respective literature. A general distinction of models can be based on whether power is directly or indirectly modelled, e.g. by mapping certain variables like measured wind speed on power via a distinct transformation. Our work contributes to direct power modelling of time series for single WTs. Most models aim at finding a precise point forecast of time series for a certain time scale and horizon \cite{modelreview2}. In contrast to that, other effort is put into modelling different properties related to power output such as the power curve \cite{kolumban2017short, gottschall2008improve}, wind power ramps \cite{cui2018copula} or power density estimates \cite{jeon2012using}. Models also vary in terms of the time scale \cite{modelreview1} that ranges from ultra--short--term $(\mathrm{ms}$-$\mathrm{s})$\cite{yang2018ultra} to long--term forecasts (months) \cite{arima}.

The model proposed in this work aims at modelling short--term power time series without the objective to give precise point forecasts. Instead, we aim at deepening the systemic understanding of a WT due to its stochastic nature on the one hand and the impact of control mechanisms (e.g. curtailment) on the other hand. Following this motivation, the model is based on a single nonlinear stochastic differential equation that is split into two components in a Langevin fashion. Since WTs are directly coupled to the complex atmospherical dynamics of wind, it is paramount to incorporate a stochastic component that allows for a certain degree of complex diffusive behaviour. A sufficient approach to include such complex diffusive dynamics without a loss of simple applicability is given by Fractional Gaussian Noise (FGN) \cite{colorednoise}. Time series generated as FGN entail a certain degree of correlation and yield Fractional Brownian Motion when cumulated.
Thus we are able to reflect on the results from the correlation analysis from a model perspective.
The second component of our model takes the impact of control mechanisms on power output into account. We address this feature in two steps: a deterministic component in the differential equation accounts for the characteristical shape of the PDF as a first mechanism to focus the power values around the respective fixed points. The second step incorporates control mechanisms such as curtailing in a simplified numerical fashion. Finally, we include a simple first approach to account for a time dependent seasonal drive of power output aswell.  
By constructing our model in such a way, its easily distinguishable components and parameters give way to a better understanding of how certain theoretical features affect power time series qualitatively, such as the degree of autocorrelation. It further yields the opportunity to test parametrized scenarios and may be used in large power network simulations based on simplified models of single wind turbine dynamics.

The paper is organised as follows: We present the data set and perform a cleansing procedure on it in sec. \ref{sec2} such that we can get a first impression of its fundamental characteristics afterwards. In sec. \ref{sec3}, we identify the autocorrelation of power time series via the traditional autocorrelation function and the method of DFA. The stochastic model we introduce in sec. \ref{sec4} is based on these empirical findings and will enlarge upon our understanding of how varying autocorrelations have an impact on fundamental statistical features by comparison with the data. When we present the results, we will find a sufficient agreement between the empirical and the modelled features. We summarize our results in sec. \ref{sec5}.

\section{Data Treatment and Characteristic Features}
\label{sec2}
We briefly introduce the data set in \ref{sec2.1}. The data cleansing procedure described in \ref{sec2.2} is important to focus only on a reasonable subset of the empirical data \cite{na}. Finally, we get a first glance of the most substantial characteristic features of the wind farm data in \ref{sec2.3}.

\subsection{Data Set}
\label{sec2.1}
The data set we analyse comprises time series of 30 WTs located at the German offshore windfarm  RIFFGAT. Several observables are measured via a SCADA (Supervisory Control and Data Acquisition) system, however we will focus only on the active power output of the WTs. The respective time series cover a total period of one year between 01/03/2014 and 28/02/2015. All analysis is carried out on ten minute average values calculated from data points measured with 1 Hz frequency. Thus the timestamp precision is limited to ten minutes and we obtain 52560 values in total.

\subsection{Data Cleansing}
\label{sec2.2}
In the following we briefly present the applied data cleansing procedure. We detect outliers and ensure that we only consider a reasonable subset of the initially measured data. Every data value we consider to be erroreneous will be set to \textsc{NA} (\textit{not available}) and is not included in any upcoming analysis.
In a first step we ensure that there are no redundant timestamps in any time series and look for consecutive identical values. Since one ten minute average value is based on 600 measured values, it can be rated as a highly unlikely case that two consecutive 10--minute averages are identical with a five digit precision in the data. This could only be conceivable if e.g. constant rated power is generated for ten minutes without any variation. To rule out this case, we also took the maximum and minimum values for the respective ten minute intervals into account.

Apart from the ten minute average values, we inspect the respective standard deviations. Any data value with a vanishing ten minute standard deviation is set to NA. In a last step we analyse if there are unreasonable changes of consecutive power values which we will call power increments. 
We consider each power increment to be unreasonably high (regardless of its direction) if it meets both of the following criteria: the power increment $\Xi_+ = \left. \left(P(t+1) - P(t)\right)\middle/ P_+\right.$ relative to the so called rated power output $P_+$ of the WT is higher than a certain threshold $\Xi_0$ and the respective minimum power $P_{\mathrm{min}}$ in a ten minute time interval is higher than the maximum power $P_{\mathrm{max}}$ in the previous time interval by a certain factor $q$: $P_{\mathrm{min}} \, > \,qP_{\mathrm{max}}$ which yields unphysical data.
We choose the threshold value $\Xi_0$ and the factor $q$ in a way that limits extreme power increments to a typical value found in the respective literature \cite{wan2004wind}. This yields $\Xi_0 = 0.67$ and $q = 0.99$.
The higher these values are chosen, the more unphysical ramps are still kept in the data which results in a higher number of strong jumps between low and high power generation. If we apply a too strict choice, some of the true strong ramps that resemble intermittent fluctuations are spuriously eliminated, also biasing results for temporal correlations. After applying the stated cleansing steps, we obtain $9.58\%$ of NA values in the data. We will only consider pairs of values containing no NA value in every calculation of correlations between time series.

\subsection{Bimodality and Power Increments}
\label{sec2.3}
To achieve a sufficient understanding of wind power data, some basic facts about the conversion of wind speed $u$ into active power output $P$ and the control of WTs have to be outlined \cite{manwell2010wind, windbook}. Although an increase in wind speed obviously leads to a higher gain of generated power in general, WTs only run within a certain operating range. This limitation is due to the finite performance of power generators. In fact, the operation of WTs is limited by a lower cut-in value $u_-$ and an upper cut-off value $u_+$ of wind speed $u$. Below $u_-$, there is simply not enough wind energy for an economic use of the turbine so that $P = P_- (= 0\,\mathrm{kW}\,)$. When $u$ exceeds $u_+$, power is controlled to remain constant at the rated power output $P = P_+ (= 3600\,\mathrm{kW}\,)$. For even higher wind speeds, it becomes essential to avoid mechanical damage and the WTs are turned down by a continuous adjustment of the rotor blades. Within $ u_- \le u \le u_+\,$, the generated power of an ideal WT increases proportionally to $u^3$. For this work the most important conclusion to draw from these control mechanisms is that power time series have to be at least in parts artificially flattened. We expect power values to be constant at $P = P_-$ or $P = P_+$ for certain time periods.

\begin{figure}[!h]
\begin{subfigure}[t]{.5\textwidth}
\includegraphics[width=8.5cm]{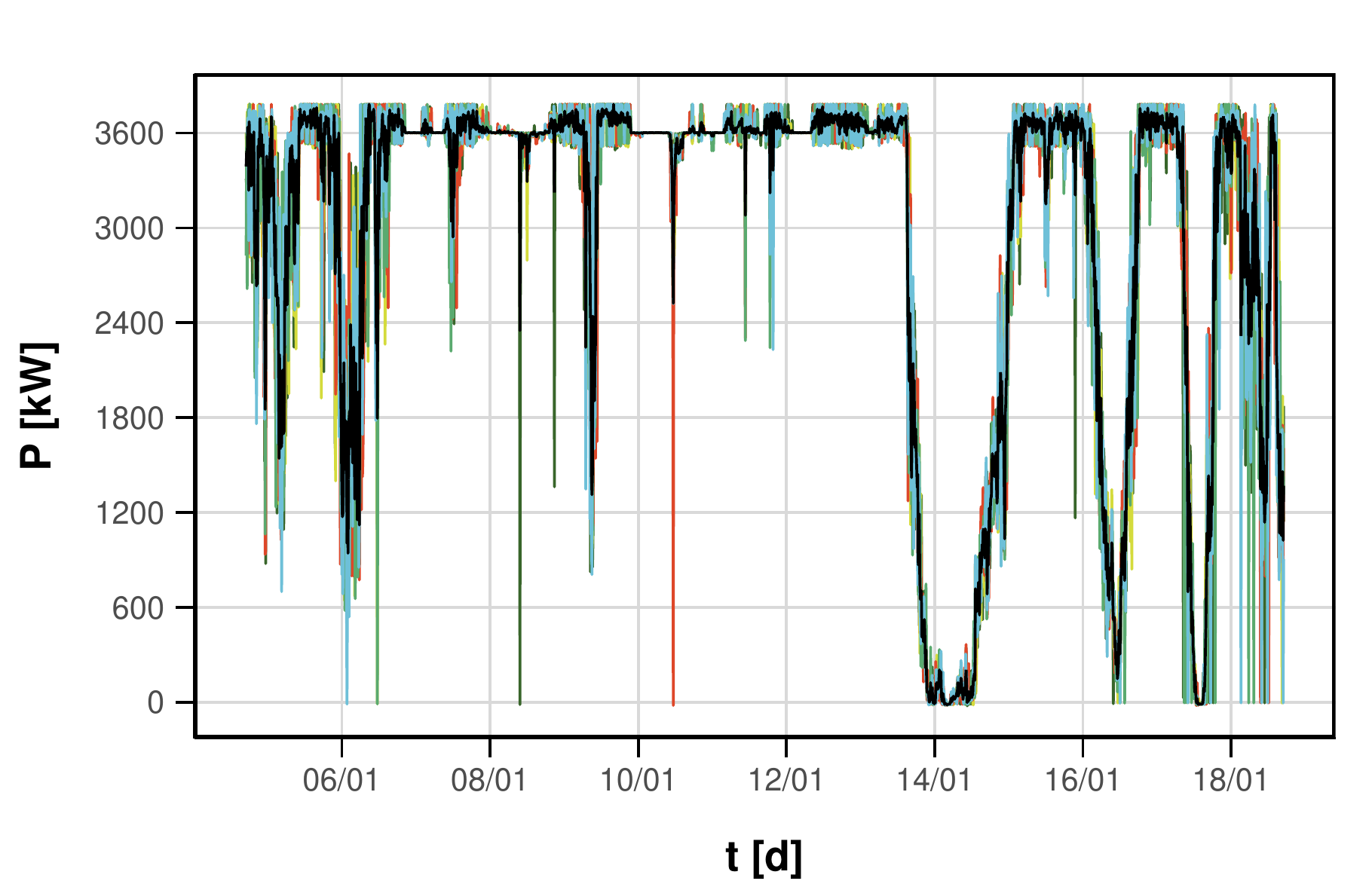}
\caption{Power time series.}
\label{fig1a}
\end{subfigure}
\begin{subfigure}[t]{.5\textwidth}
\includegraphics[width=8.5cm]{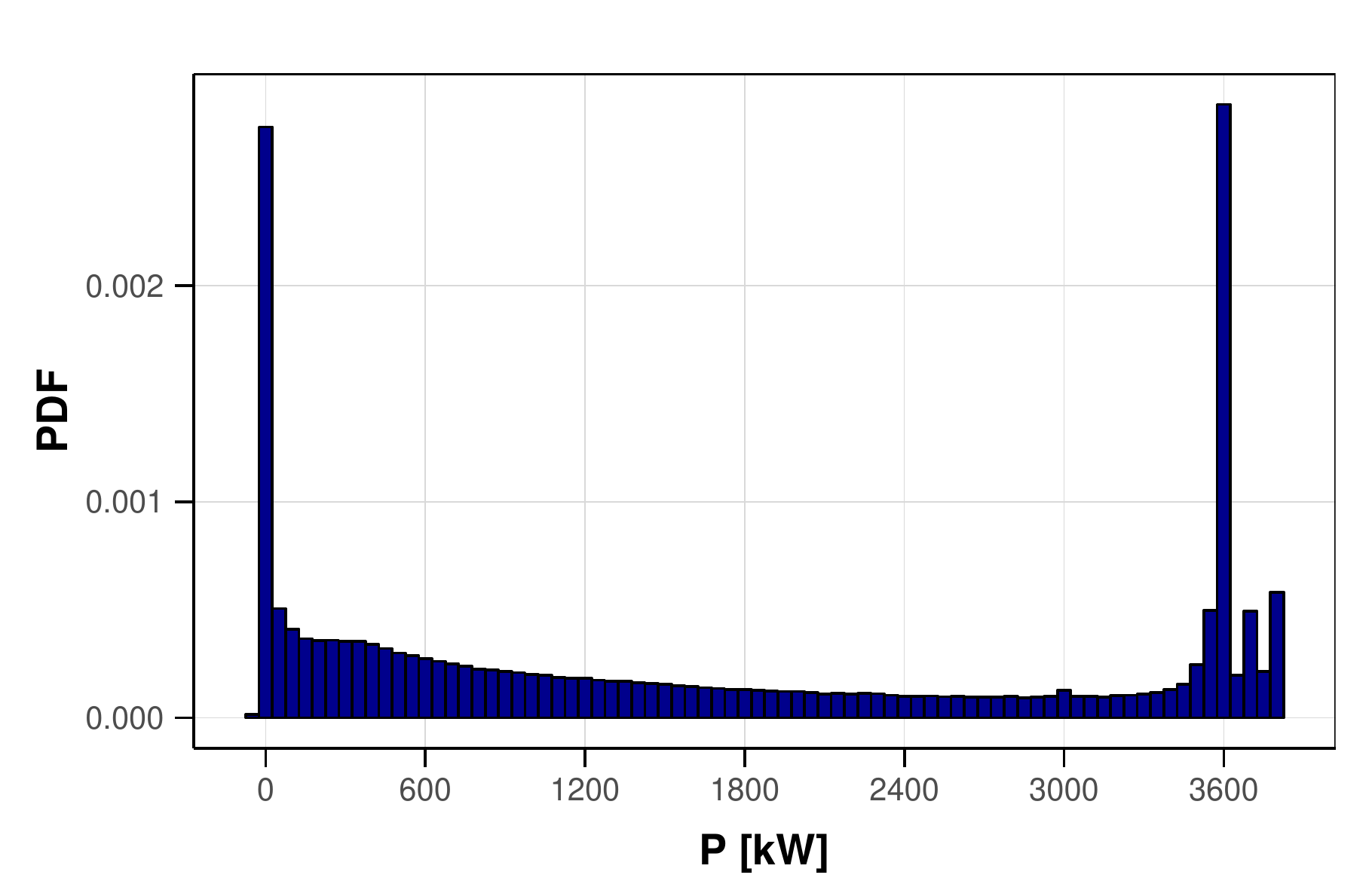}
\caption{Empirical PDF as histogram of all power values.}
\label{fig1b}
\end{subfigure}
\caption{Exemplary power time series and average power output (black) covering two weeks. The respective PDF plotted as a histogram is shown for all WTs covering the entire time period.}
\label{fig1}
\end{figure}

This manifests itself in fig.\ref{fig1a} as a striking bimodal pattern in the five displayed time series and a striking bimodality of the empirical PDF in fig.\ref{fig1b}, visualized as a histogram. Most power values are concentrated around zero power generation $P_-$ and active power output $P_+$. In fact, the intervals $50\,\mathrm{kW} < P < 200\,\mathrm{kW}$ and $3550\,\mathrm{kW} < P < 3800\,\mathrm{kW}$ around the two peaks sum up to $41.62\%$ of all values. This feature also governs the dynamics observed from the time series and is observed in several different works \cite{scholzmcmc,jeon2012using,powerdistrexmpl4, betadistr}. Nevertheless, other analyses \cite{powerdistrexmpl1,powerdistrexmpl2} also show unimodal distributions around $P_-$ or flatter, less concentrated distributions. While the first observation is related to the different efficiencies of WTs, the latter is mostly found for aggregated power of several wind farms where WTs with different rated power outputs are combined. Despite of this bimodal shape of the PDF, also values exceeding the rated power can be observed. Finally, strong downward ramps to zero power output can be observed for some of the WTs.

\FloatBarrier

\begin{figure}[!h]
\begin{subfigure}[t]{.5\textwidth}
\includegraphics[width=8.5cm]{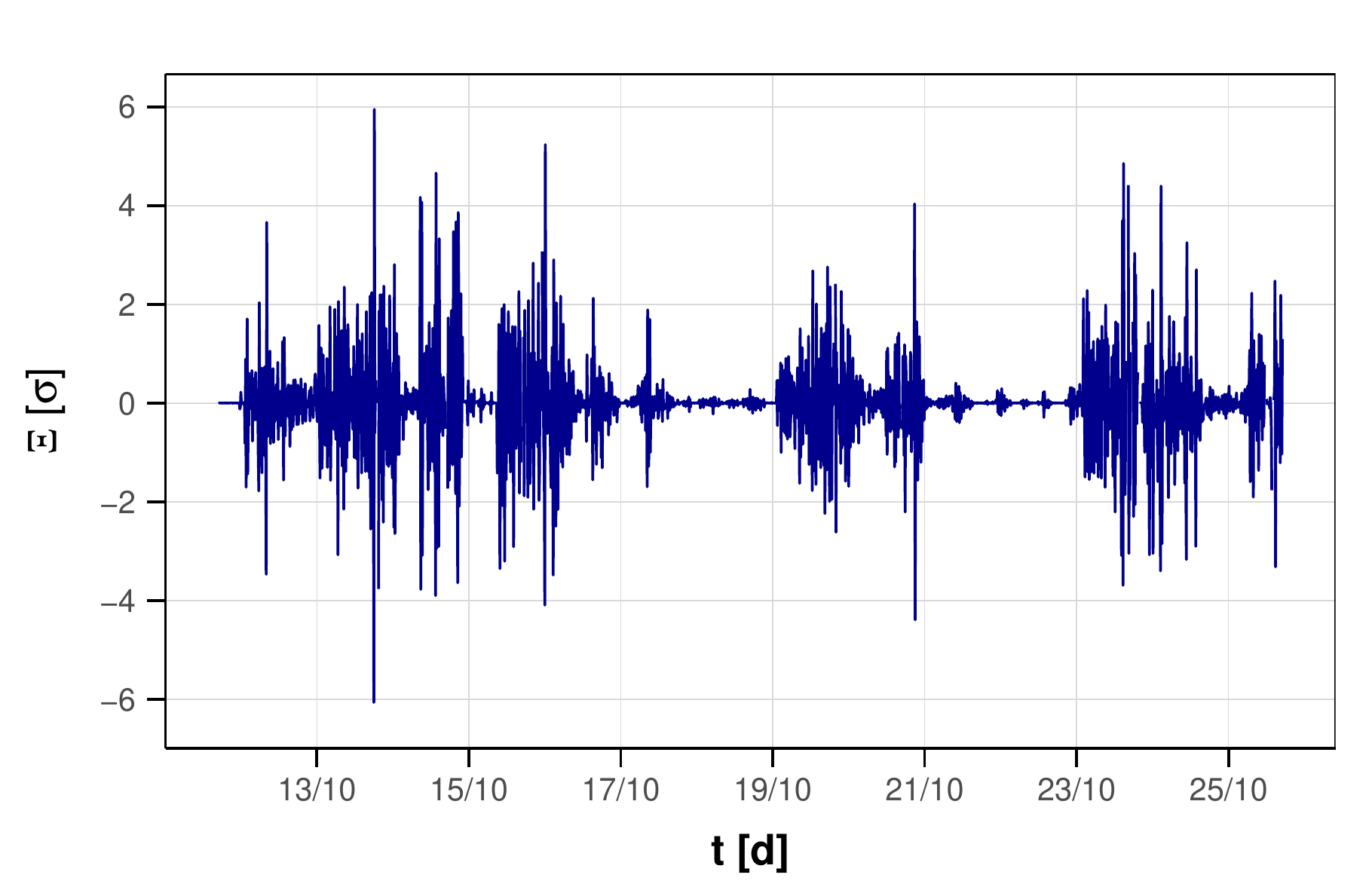}
\caption{Power increment time series.}
\label{fig2a}
\end{subfigure}
\begin{subfigure}[t]{.5\textwidth}
\includegraphics[width=8.5cm]{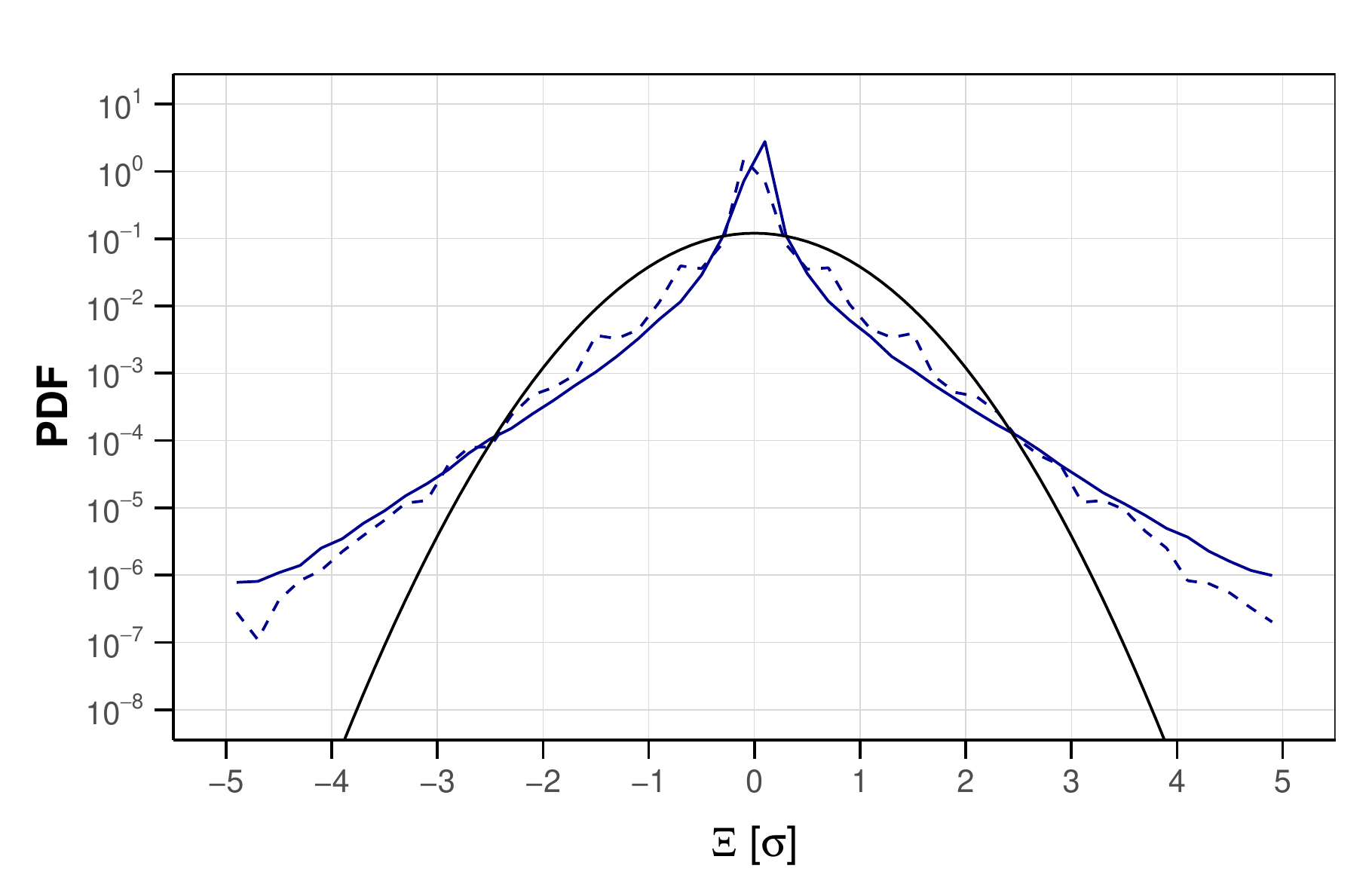}
\caption{Single--logarithmic histogram of all power increments.}
\label{fig2b}
\end{subfigure}
\caption{Exemplary power increment time series covering two weeks and respective single--logarithmic histogram for all WTs covering the entire time period, compared to a Gaussian distribution. The dashed line includes power increments of the aggregated wind farm.}
\label{fig2}
\end{figure}

We will characterize such power ramps by the increments $\Xi_k(t) = P_k(t+\Delta t) - P_k(t)$ which are a fundamental property for the understanding and management of wind farms.
We define them as the standardized, dimensionless differences
\begin{align}
\tilde{\Xi}_k(t) \, = \, \frac{\Xi_k(t) \ - \ \frac{1}{T}\sum_{t=1}^T \Xi_k(t)}{\sqrt{\frac{1}{T}\sum_{t=1}^T \left(\Xi_k(t) - \mu_k\right)^2}}
\label{eq1}
\end{align}
For our data, we only analyse $\Delta t = 10\,\mathrm{min}$. A visual inspection of fig.\ref{fig2a} shows an example of this property for one WT in a time period of two weeks.
Apparently, increments of similar size cluster in time, indicating some degree of correlation. This generally indicates that simple random walk models do not yield a sufficient description that captures the complex temporal correlations of time series. Power seems to fluctuate symmetrically but clearly in a non--Gaussian fashion as it can be seen in the PDF of all $\Xi_k(t)$ for the total time period in fig.\ref{fig2b}. Here we compare the empirical distribution to a Gaussian (black) with the same mean value and standard deviation. For the aggregated wind farm, the strong fluctuations appear slightly dampened. These findings are in accordance with results in the literature for increments on even shorter time scales\cite{anvari2016short}. 

\FloatBarrier
If the PDF deviates from Gaussain statistics we speak of intermittent behavior after Kolmogorov 1962, expressed by a heavy tailed distribution and multifractal statistics. Note this turbulent intermittency term is not the same as the alternative denotation of intermittency, i.e. nonstationary time series switching between different flow states like between laminar and turbulent flows.

\section{Autocorrelation}
\label{sec3}

Both the bimodality of power, persisting at values around zero and rated power output, and the complex dynamics of power increments suggest that an analysis of correlations can be fruitful.
As a first step, we display linear autocorrelations for power time series $P_k(t)$.

To this extent, we use the Pearson Correlation with a time delay $\tau$ defined by

\begin{align}
\Theta(\tau) \, = \,\left. \left[\langle X(t)X(t + \tau)\rangle_t \, - \, \left(\langle X(t) \rangle_t\right)^2\right] \middle/ \langle (X(t))^2 \rangle \right.
\label{eq2}
\end{align}

\begin{figure}
\begin{subfigure}[t]{.5\textwidth}
\includegraphics[width=8.5cm]{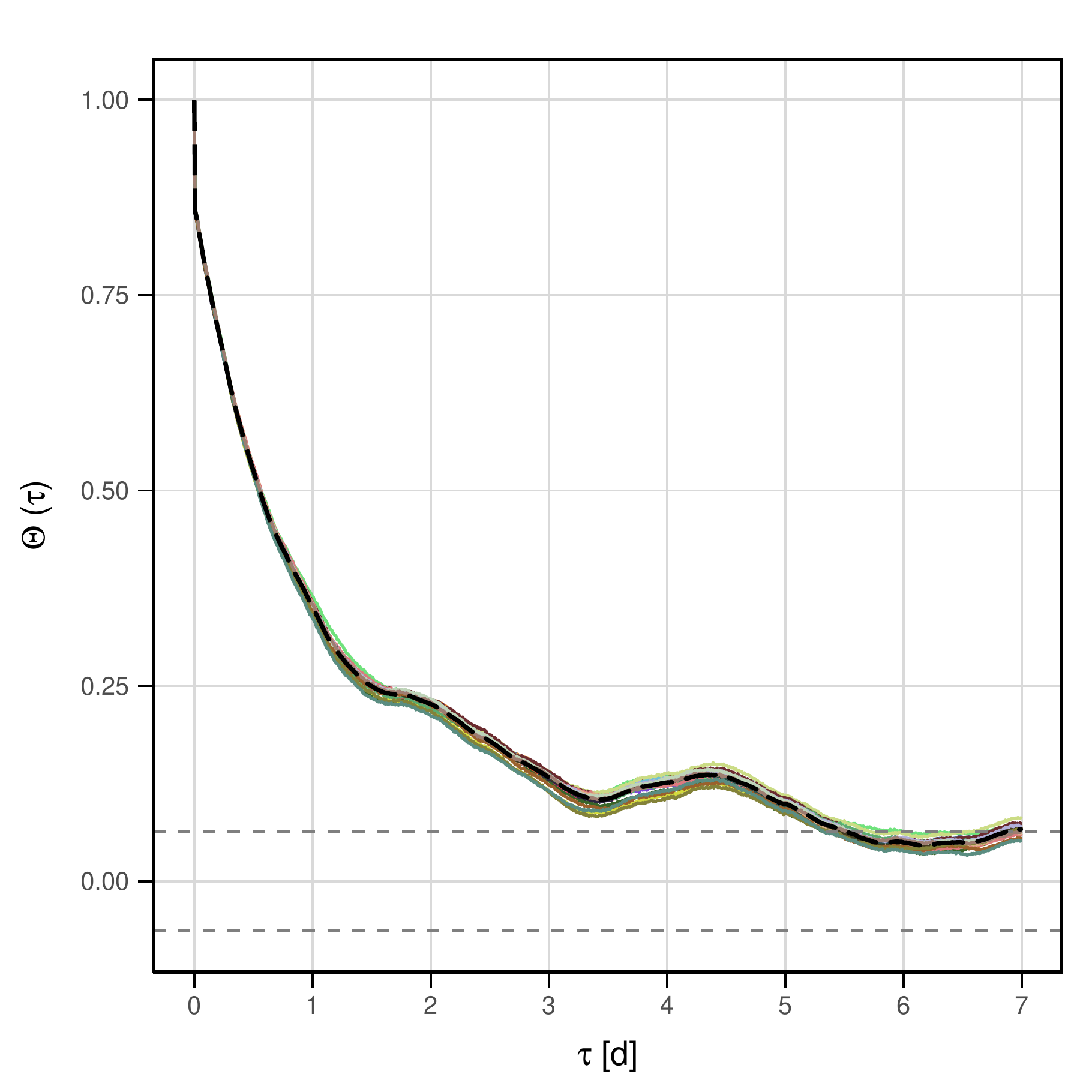}
\caption{Linear autocorrelations of all $P_k(t)\,$.}
\label{fig3a}
\end{subfigure}
\begin{subfigure}[t]{.5\textwidth}
\includegraphics[width=8.5cm]{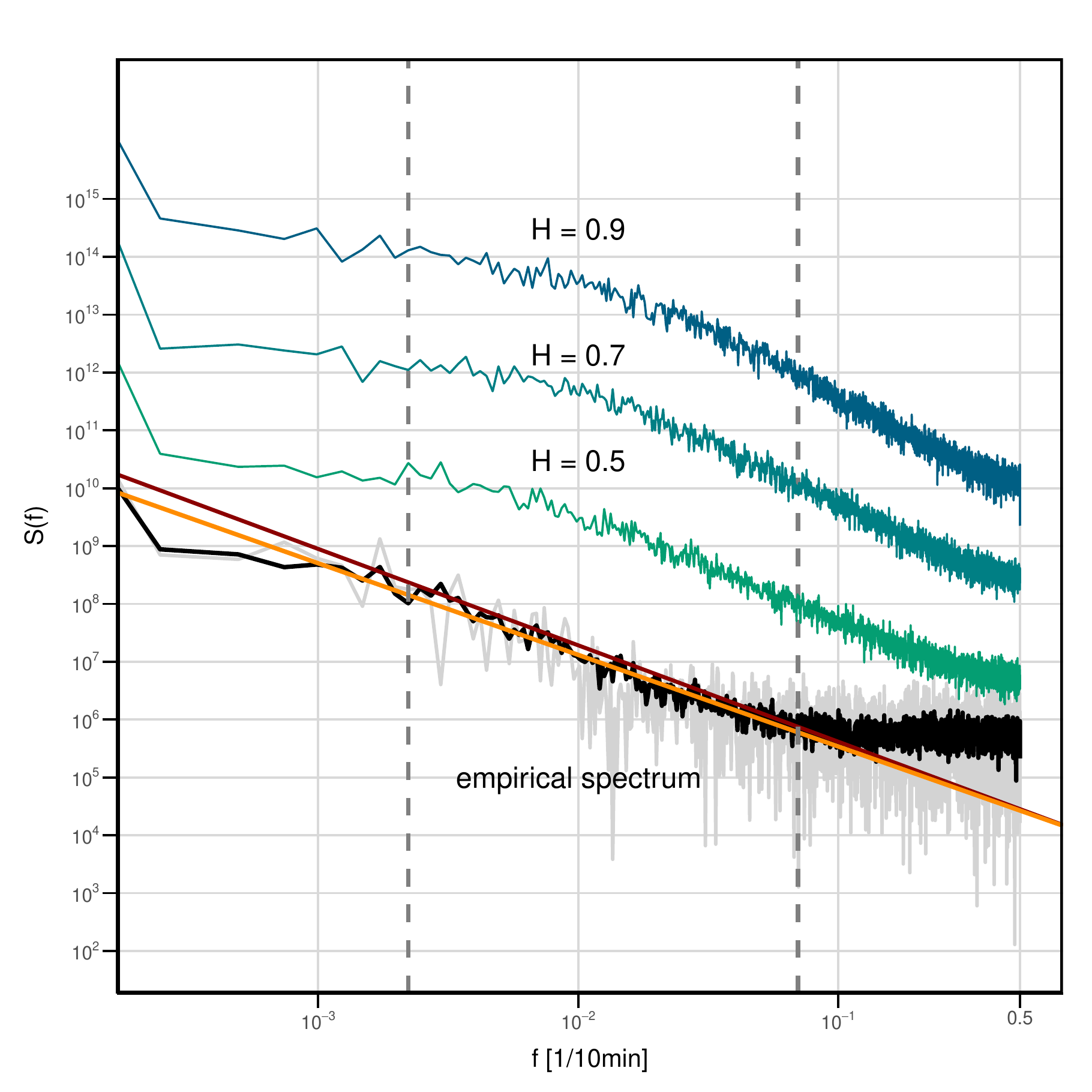}
\caption{Fourier power spectrum of exemplary $P_k(t)\,$.}
\label{fig3b}
\end{subfigure}
\caption{Autocorrelation and Fourier spectrum of power time series.
(a): $\Theta(\tau)$ of all 30 WTs power time series for $\tau\leq 1\,\mathrm{week}\,$. The average autocorrelation is drawn as a black dashed line. The dashed horizontal lines represent a level of significance.
(b): Fourier power spectrum of an exemplary power time series with log-log-scale. The time series was split into ten equally sized segments for which the power spectra were computed (example: gray) and averaged subsequently (black). The orange line shows the linear regression performed within the fitted region (dashed lines), the red line displays a scaling with $\beta = 5/3$. The three additional spectra are equivalently computed for the optimized model time series (see. fig.\ref{fig8}) with the respective Hurst exponent and shifted for better visibility.}
\label{fig3}
\end{figure}

Quantifying autocorrelations of power time series yields valuable information on how power can be generally modelled (sec. \ref{sec4}). In this paper, we only consider autocorrelation of power time series. With $\Theta(\tau)$ as a correlation measure we only account for linear dependence. Moreover, it is sensible to outliers and only gives sufficient information for time series with finite variance. Figure \ref{fig3a} shows $\Theta(\tau)$ for power time series of all 30 WTs. It is shown up to a lag of seven days which is approximately the point where they drop below a significant level (dashed horizontal lines). As such we use a simple surrogate approach and shuffle the time series, eliminating temporal information but preserving the PDF. We identify a (constant) confidence band by calculating its width as $2\sigma$ of the autocorrelation after the shuffeling process, averaged over all time series. The autocorrelation of all $P_k(t)$ slowly decreases over three orders of magnitude and thus indicates long--range dependence. Nevertheless, $\Theta(\tau)$ does not show a typical power law decay but runs through several local maxima, reflecting the inherent nonstationarity. In \cite{mcmc} an explanation for this observation is provided that corroborates our results: the autocorrelation of wind speed decreases with slightly visible maxima due to weather related seasonalities. Since power is closely coupled to wind speed and persists at an almost constant level for values around $u_-$ and $u_+$, the local maxima are not only sustained but amplified. The displayed autocorrelations do finally not show significant differences between single WTs even though it is known that relative positions of WTs play a role for power generation, e.g. through wind shear effect  \cite{wen2017power}. Such differences could potentially become visible in lagged cross--correlations between WTs of varying relative position which is not within the scope of this work though.

\FloatBarrier
The scaling behaviour of power time series is expected to be related to that of wind time series. For the latter, it is well known that the Fourier power spectrum obtained yields a power law decay $E(f) \propto f^{-\beta}$ with $\beta = \left. 5 \middle/ 3\right.$ estimated through linear regression in the log-log plot. This finding is in accordance with results one obtains from Kolmogorov's turbulence model.
A simple way to confirm that similar scaling can be found for the power time series, we estimate $\beta$ from a respective Fourier power spectrum in fig.\ref{fig3b}. To this extent, we compute a smoothed spectrum (black) by splitting the full time series $P(t)$ of an exemplary WT into ten chunks and average over spectral power values obtained for each of the single segments (example in gray).
The linear MLE regression (orange) yields $\beta = 1.58 \approx \left. 5 \middle/ 3\right.$ within a linear region of sufficient width and thus matches both the expected universal scaling (red) and results from similar data well \cite{medina2015multiscale}.

The presence of nonstationarity leads to biased or spurious detection of autocorrelations \cite{lrdpitfalls}. To uncover autocorrelations in presence of nonstationary we apply DFA \cite{dfadetrmethods} to the power time series $P_k(t)$.
The main idea of DFA is to eliminate nonstationarity in a generalized fashion by subtracting polynomial trends and consequently focusing on correlations that are present in the remaining noise. The method aims at calculating the Hurst exponent $H$, introduced by Hurst in 1951 \cite{hurstoriginal}. We distinguish between persistent ($0.5 < H < 1$) and antipersistent ($0 < H < 0.5$) behaviour of a time series. The special case $H = 0.5$ yields diffusive behaviour i.e. uncorrelated white noise with $\langle X^H(t)X^H(t')\rangle = 0$ for the time series $X(t)$ at arbitrary times $t$ and $t'$. Cumulated white noise entails Brownian Motion. 
The persistent regime $H > 0.5$ results in super--diffusive dynamics and long--range dependence of the increments $X(t)$. Antipersistence yields the opposite case, i.e. a time series changes its direction more frequently than a diffusive time series. The respective extension of a white noise process is called Fractional Gaussian Noise (FGN) and will be adressed later.

We now briefly sketch the method of DFA and refer to \cite{dfadetrmethods} for a more detailed description. As a first step we calculate the integrated, mean adjusted power time series. A time series of equal length is obtained that is splitted into $\,N_s = \lfloor T/s \rfloor\,$ disjunct subsets of equal length $s$. The brackets round the value of $T/s$ down to an integer value. We repeat this step with the reversed time series to incorporate all subsets. 
Subsequently, a polynomial quadratic detrending via MLE--regression (Maximum Likelihood Estimation) is applied for all subsets respectively. We then calculate the standard deviation $F_{\nu}(s)$ of all detrended subsets and from that derive the average standard deviation. Repeating this calculation for different $s$, we obtain the fluctuation function

\begin{align}
F(s) \ = \  \sqrt{\frac{1}{2N_s} \sum\limits_{\nu=1}^{2N_s} F_{\nu}^2(s)}
\label{eq3}
\end{align}

that represents the scale dependent fluctuation strength. We estimate the scaling exponent $\alpha$ from $F(s) \propto s^{\alpha}$ empirically via linear regression in a double--logarithmic plot since we expect $F(s)$ to increase as a power law. For stationary time series (e.g. FGN \cite{taqqu1995estimators}), we can identify $\alpha$ with the Hurst exponent $H$. In the more general case frequently encountered for real data, even for the detrended time series some nonstationarity remains present. In this case, the Hurst exponent can only be estimated as $H \approx \alpha - 1$ which only strictly holds for a Fractional Brownian Motion (FBM) process. This distinction is sometimes not pointed out distinctively in the literature \cite{crossoverwindspeed}. In general, $\alpha$ is also linked to the Fourier scaling exponent $\beta$ via $\beta = 2\alpha - 1 = 1 + 2H$, enabling us to assess the consistency of our results.

\begin{figure}[!h]
\centering
\includegraphics[width=10.5cm]{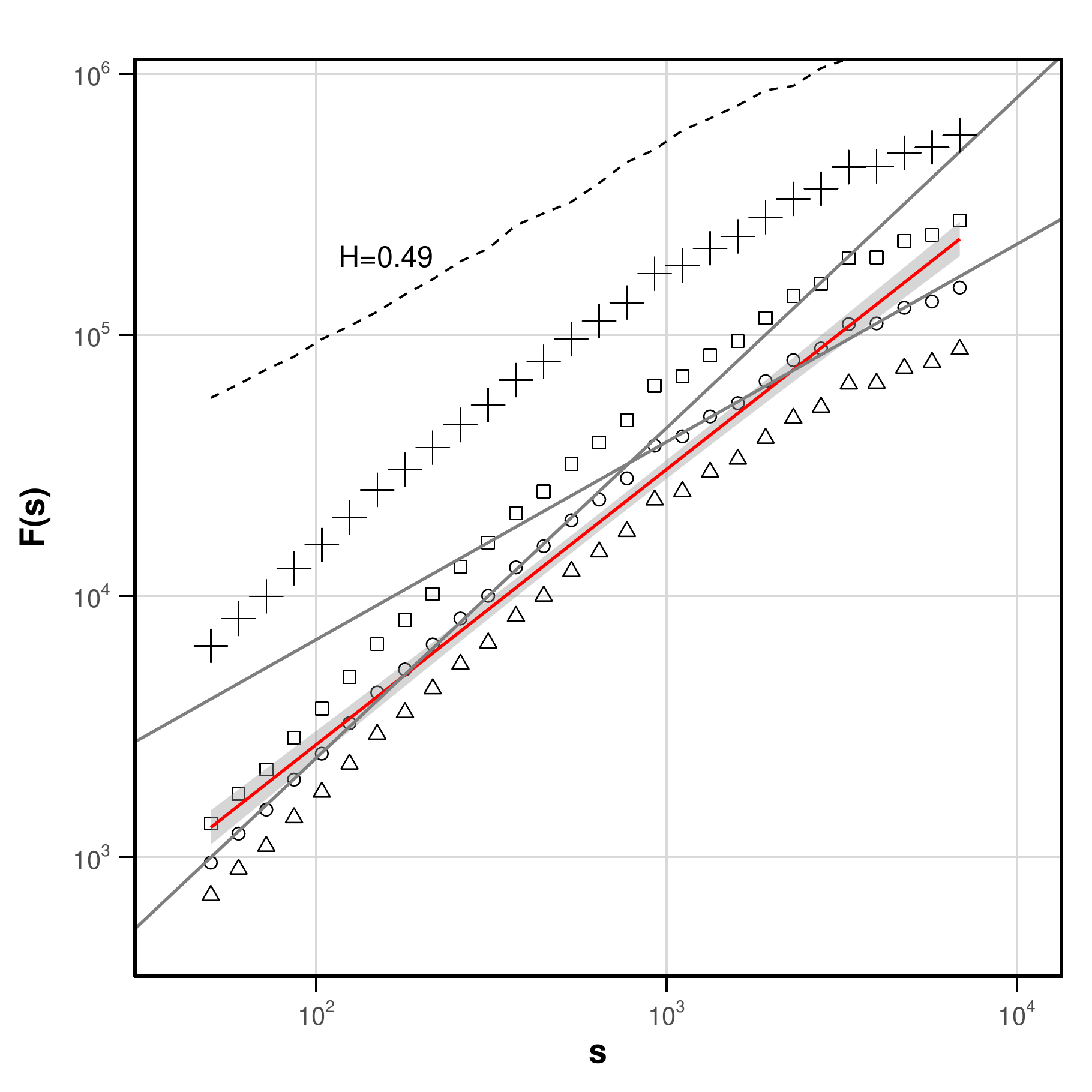}
\caption{Fluctuation function $F(s)$ for three exemplary WTs and aggregated WF (crosses). $s$ is given in units of 10 min. The curves are shifted in $y$--direction for better visibility. The red line displays the MLE--regression with shaded error region. The gray lines visualize the crossover. The dashed line shows $F(s)$ for a surrogate of randomly shuffled data.}
\label{fig4}
\end{figure}

For applications, $s$ must not be chosen too small for a significant outcome. The DFA method is known to cause misleading finite size effects with respect to short time scales $s$ and the applied detrending generally needs to be regarded critically \cite{dfarevisiting}. Another typical issue are values at the limits or outside of the range $0 < \alpha < 1$. For $\alpha > 1$, we have to consider the time series as an integrated process with more complex nonstationarities ($H = 1.5$ equals Brownian Motion) \cite{dfaconsistency,anomaldiff}. Apart from that, the resulting increase of $F(s)$ does not have to be completely linear but can contain crossovers with different slopes \cite{dfaheart}. In many cases, these crossovers are meaningful and uncover different scaling regions that entail different correlations \cite{kantelhardtdfa}.
As we see in fig.\ref{fig4}, the displayed fluctuation functions follow power laws with similar scaling exponents $\alpha$. We checked that the displayed results are robust to different orders of polynomial detrending. Three shifted curves are shown for single WTs and one for the aggregated windfarm. A linear increase with one clearly visible crossover $s_{\mathrm{c}}$ at a time scale of approximately three days can be identified for all of the curves. The resulting scaling exponents are $\alpha = 1.33$ for $s\leq s_{\mathrm{c}}$ and $\alpha = 0.80$ for $s > s_{\mathrm{c}}\,$, averaged over all obtained $\alpha$ for the different turbines. The values are almost identical for the aggregated windfarm.
The dashed line emphasizes that all found results in fact give meaningful information about the correlations and not only the distribution. The line refers to a stationary random surrogate time series we obtained with the same procedure explained above and applying DFA afterwards. A value of $H_{\mathrm{surr}} = 0.49$ reflects diffusive behaviour.  
Consequently, for short time scales $s \le s_{\mathrm{c}}$ power time series have to be regarded as a highly nonstationary stochastic process with trends that can not be sufficiently eliminated by DFA. If we suppose the validity of $H \approx \alpha - 1$ for $s \le s_{\mathrm{c}}$, we obtain $H \approx 0.33$ matching $H \approx 0.5(\beta - 1) = 0.29$ derived from the Fourier spectrum in fig.\ref{fig3b}. 
For $s > s_{\mathrm{c}}\,$, the scaling exponent $\alpha = 0.80$ yields persistence and a Hurst exponent $H \approx -0.20$. The latter finding is consistent with the displayed autocorrelation functions and values of $\alpha$ found in the literature \cite{lrdmfwindspeed, multifractalpower, medina2015multiscale}.

The same type of crossover behaviour is also found for hourly wind speed data at geographically different sites \cite{crossoverwindspeed}. The authors conjecture that this scaling behaviour might arise from the different scales of weather patterns which would manifest itself in a multifractal spectrum of different scaling exponents. Yet, the crossover is not critically reflected on, even though misleading crossovers in DFA are known to appear with several known causes. In \cite{kantelhardtdfa} it is suggested to test whether different orders of polynomial detrending change the crossover position which we ensured not to occur. Furthermore, the number of NA--values does not have a significant impact on our results as supposed in \cite{dfana}.
Although the scaling law of $F(s)$ is still valid for $\alpha>1$ \cite{lovsletten2017consistency}, this special case hints that unidentified trends remain after the detrending procedure. Such trends are also known as a source of erroneous crossovers \cite{dfatrends, dfanonstationary} which we can not fully rule out. If such trends were the cause, similar trends would also be present in the wind speed data in \cite{crossoverwindspeed} though.
The consequences of such a crossover still seem to be of potential importance for possible modelling approaches which will be adressed in sec. \ref{sec4}.
.
\\

\FloatBarrier

\section{Stochastic Model}
\label{sec4}

The identified features of single WT power time series motivate a model approach that could be calibrated with empirical data. We now put forward such a time series modelling approach that has the potential to be used in a larger simulation, aiming at a multi-scale model of wind power generation. In sec. \ref{sec4.1}, we introduce our model approach. We compare our model results to the empirical data in sec. \ref{sec4.2} to gain first qualitative insights into the scope of the model.

\subsection{Model Approach}
\label{sec4.1}

Numerous approaches of power time series $P(t)$ simulations are employed in the literature. A major fraction tends to simulate wind farm power time series only and does not concentrate on single WTs as we intend to do \cite{olauson2017simulating, fertig2019simulating, trombe2012general, modelreview1}. Another central distinction is the aim of the model. Our reduced-form approach does not intend to reproduce temporally ordered forecasts but only statistical features such as distributions and long--term averages. Several models in the literature manage to give precise forecasts of different time horizons of $P(t)$ using black-box models with a high number of parameters or simple regression parameters that often lack a comprehensive contextual meaning. Typical model approaches in this context are Markov processes \cite{mcmcsecondlag, mcmc, markovintro} and ARIMA (Autoregressive integrated moving average) processes \cite{arima} which both focus on the autocorrelation of $P(t)$. Further approaches are based on nonlinear models \cite{chaos}, stochastic models \cite{scholzmcmc} and stochastic drift--diffusion--models \cite{milan}. In contrast to the stochastic process approach in \cite{milan} which aims to model the stochastic wind speed / wind power relation in seconds by the estimators of Kramers--Moyal coefficients, we aim here to achieve a stochastic modelling of the power output based on 10 minute values.

We will try to introduce our model parameters with a comprehensible meaning. Furthermore, our model equation itself is not a regression formula but is based on our qualitative understanding of power time series. The simulations are carried out on a 10--minute time scale. 
The essence of our model comes down to one fundamental stochastic differential equation (SDE) based on central statistical features we identified in sec. \ref{sec2} and sec. \ref{sec3}. We do not model NA-values separately and aim at simulating the raw uncleansed data that is directly obtained from the SCADA system \cite{forecastna}. For the sake of simplicity all model-related equations are formulated in a notation for continuous systems, yet baring in mind that we are dealing with discrete data. We now briefly introduce the stochastic model equation and summarize the related mathematical concepts.
In our modelling framework, we will regard power generation as an autocorrelated stochastic process with a deterministic and a stochastic component. These drift--diffusion--types of models are often expressed by a Langevin equation of the general form

\begin{align}
\dfrac{\mathrm{d}P}{\mathrm{d}t}(t) \ = \ -k\, \dfrac{\mathrm{d}V(P)}{\mathrm{d}P} \, + \, D\, \xi_H(t) \qquad \quad .
\label{eq4}
\end{align}

Here, $P(t)$ denotes the power time series. The first term equals the deterministic drift component with a drift parameter $k$ and a potential function $V(P)$. The second component incorporates stochastic fluctuations $\xi_H(t)$ with a constant diffusion strength $D$. The index $H$ denounces the Hurst exponent which hints at the fact that we can include arbitrary power law-like correlations into our model. Thus, $\xi_H(t)$ itself is a solution to a simple FGN--stochastic process \cite{hurstandfgn} with the defining property

\begin{align}
\langle \xi_H(t)\xi_H(t')\rangle \ = \ \frac{1}{2}\left(|t|^{2H} \, + \, |t'|^{2H} \, - \, |t - t'|^{2H} \right) \, .
\label{eq5}
\end{align}

In this way, we can transfer our empirical knowledge about the autocorrelation of $P(t)$ from sec. \ref{sec3} to the model approach. Furthermore, we choose the deterministic potential function $V(P)$ in a way that focuses on the time series dynamics around the fundamental fixed points. Since we have observed that power generation mostly concentrates around zero power output $P_-$ and rated power $P_+$, we choose a bimodal approach with the double-well potential function

\begin{align}
V(P) \ = \ \frac{1}{4a^4}(P-P_0)^4 \, - \, \frac{1}{2a^2}(P-P_0)^2 \, .
\label{eq6}
\end{align}

While $a$ describes the steepness of the potential, $P_0$ gives us the position at which it is centered.
Such SDEs are used in different applications in the literature and are refered to as a description of an overdamped Brownian particle in a double-well potential \cite{overdampbrown} in statistical physics. As an illustration, we can think of the underlying dynamics as a particle that would jump between the potential minima, driven by correlated fluctuations. The latter consequently introduce a characteristic time scale for the transitions between the fixed points. Often, such models are extended by a driving periodical force that entails a biased occupation of one fixed point with respect to a certain seasonality. Since power generation from wind energy follows seasonal variations, we incorporate this into our model with the simple approach
\begin{align}
F(t) \ = \ A\, \mathrm{cos}\,\omega t
\label{eq7}
\end{align}
which is added to eq.\ref{eq4} as the driving force.
We determine $\omega$ from Fourier analysis as the leading frequency within the spectrum of $P(t)$. $A$ gives us the adaptable strength of the seasonal variations. With this extension, we introduce another characteristic time scale into our approach. The interplay of both this component and the stochastic fluctuations determines the transition dynamics as described by the general phenomenon of stochastic resonance \cite{stochasticresonance}. Our approach effectively biases the bimodal PDF of modelled power towards one of its peaks which reproduces the seasonal variation of wind power generation on a rather basic level. In the parameter estimation of $\omega$, we incorporate data from a specific month to calibrate the seasonality according to the month in the data.

As we have seen in fig.\ref{fig1}, the control of the WTs results in extremely narrow peaks of the PDF of $P(t)$. This fact is caused by the intervening external control of power output (curtailment). Our model approach already succeeds to concentrate the power values around $P_-$ and $P_+$ as fixed points in a similar manner but fails to narrow the peaks down as sharply as required due to the simple analytical approach. Hence we have to incorporate the artificial flattening of time series we observe in the data into our model sufficiently. To do so, we cut off power values beyond the operating range $P_- \le P \le P_+$ by setting them to the respective constant threshold value $P_-$ or $P_+$. As we observe in the data, these limits are sometimes exceeded in the empirical time series anyway. Consequently, this procedure is only applied with a certain probability $p_{\gtrless}$. Note that the model can not be regarded as a typical Langevin--equation driven model since the correlated noise term $\xi_H(t)$ on the one hand and the artificial flattening due to curtailment on the other hand differ strongly from the standard delta--correlation of such models.
Taking all explained considerations of our model approach into account, the resulting model formula is

\begin{align}
\begin{split}
P(t) \ &= \ \begin{cases}
		\tilde{P}(t) \qquad \quad &  \text{if} \quad P_- \le  P(t) \le P_+ \\
		P_{\pm} \qquad \quad &  \text{if} \ \tilde{P}(t) \gtrless P_{\pm} \quad \text{with} \quad p=1-p_{\gtrless} \\
		P_{\pm} + z \qquad \quad &  \text{if} \ \tilde{P}(t) \gtrless P_{\pm} \quad \text{with} \quad p_{\gtrless}
		\end{cases}
		\\\\
\dfrac{\mathrm{d}\tilde{P}(t)}{\mathrm{d}t} \ &= \ \left[ -\left( \frac{\tilde{P}(t) - P_0}{a}\right)^3 + \frac{\tilde{P}(t) - P_0}{a} + A\, \mathrm{cos}\,\omega t \right] \, + \,  D \xi_H(t)
\end{split}
\label{eq8}
\end{align}

with the respective probabilities $p_{\gtrless}$ for having a power value beyond the operation range and a $\mathcal{N}(0,\sigma)\,$--distributed random number $z$. The model includes ten parameters in total. A detailed description of the parameter calibration and all resulting values are given in app. \ref{app}. Six parameters ($P_0,\, \omega,\, p_{\gtrless},\, \sigma_{\gtrless}$) are calibrated on the data before a time series is to be modelled. $P_0$ determines around which power value the distribution of values should be centered. The frequency $\omega$ should include some limited degree of season-specific variation and is fixed via Fourier analysis. The remaining four parameters $p_{\gtrless}$ and $\sigma_{\gtrless}$ calibrate the curtailment and variability of power beyond the operating range. The three parameters $a\, , \, A$ and $D$ are optimized such that the model reproduces the empirical statistical features most effectively while avoiding overfitting. The Hurst exponent $H$ will be varied in sec. \ref{sec4.2} to observe how different autocorrelations affect the generated power. 

\subsection{Results}
\label{sec4.2}

We get a first glance of the model results by inspecting simulated time series. We vary the Hurst exponent $H$ from diffusive ($H=0.5$) to persistent ($H=0.7$) up to strongly persistent ($H=0.9$) dynamics to account for different degrees of correlation within the model. Thus, our approach does not incorporate the uncovered crossover behaviour that would separate between different autocorrelations below and above $s_c$ but only account for scales $s<s_c$ in the stochastic component. Yet the deterministic component results in strong ramps that are intended to resemble the observed power ramps that lead to the Hurst exponent $H>1$ for $s>s_c$.
We always compare the simulated results to only one exemplary WT since the model aims at characterizing the typical dynamics instead of a single specific WT.

\begin{figure}[!h]
\centering
\includegraphics[width=.75\textwidth]{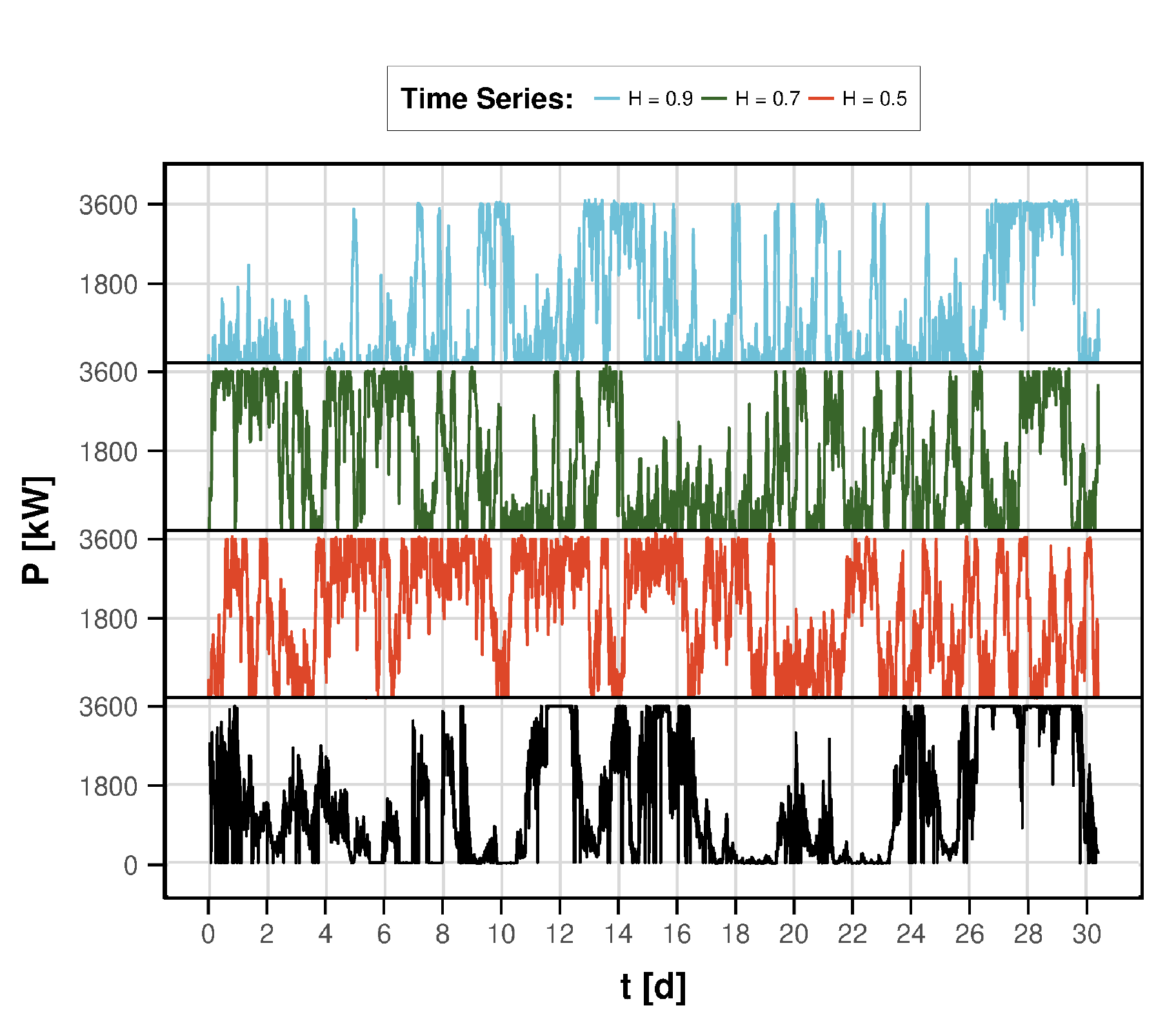}
\caption{Comparison between empirical (bottom) and simulated power time series for different Hurst exponents $H$. The black line represents a typical empirical time series for one month. The simulated time series are generated for the same month and chosen randomly.}
\label{fig5}
\end{figure}

\begin{figure}[!h]
\centering
\includegraphics[width=8cm]{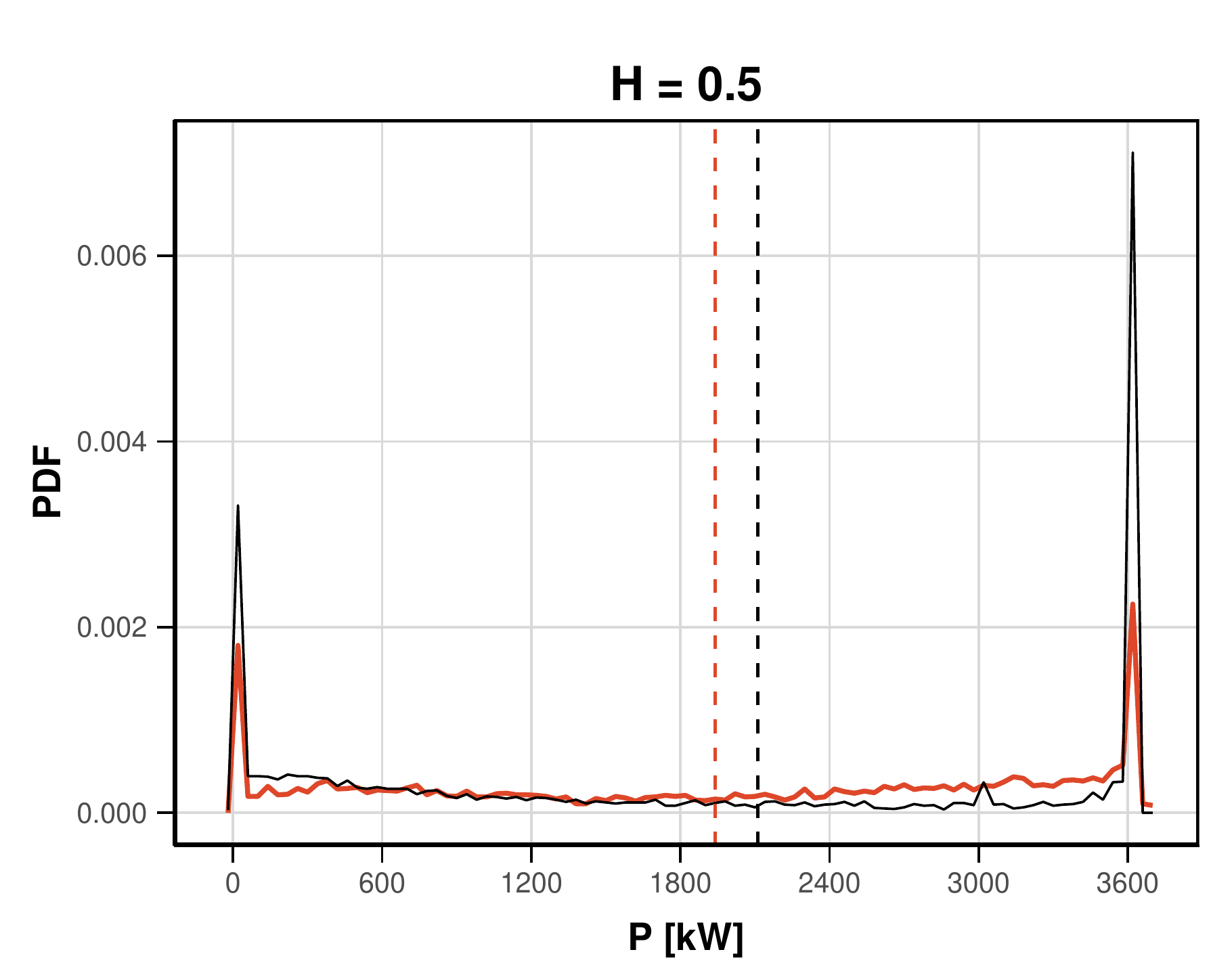}\\
\includegraphics[width=8cm]{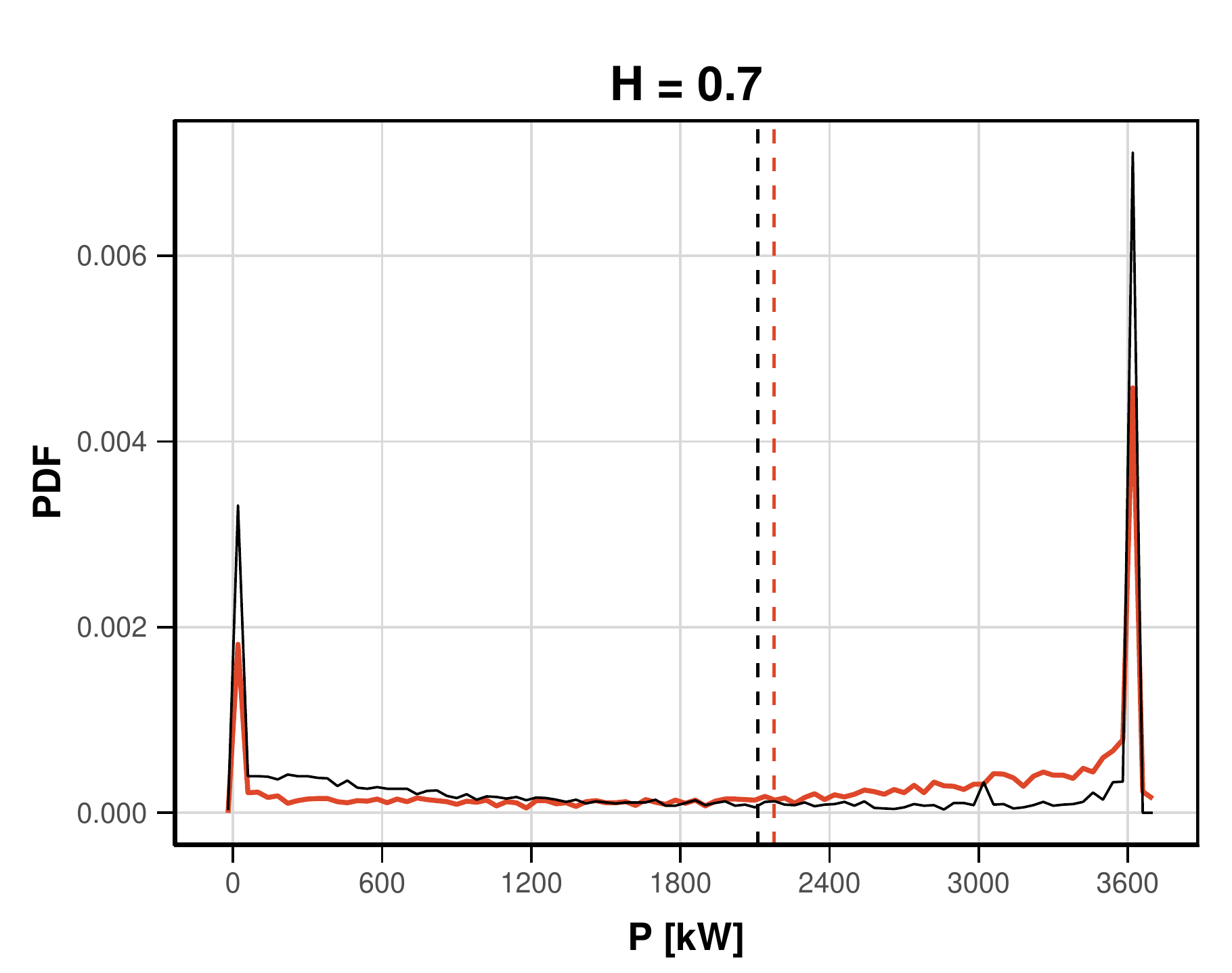}\\
\includegraphics[width=8cm]{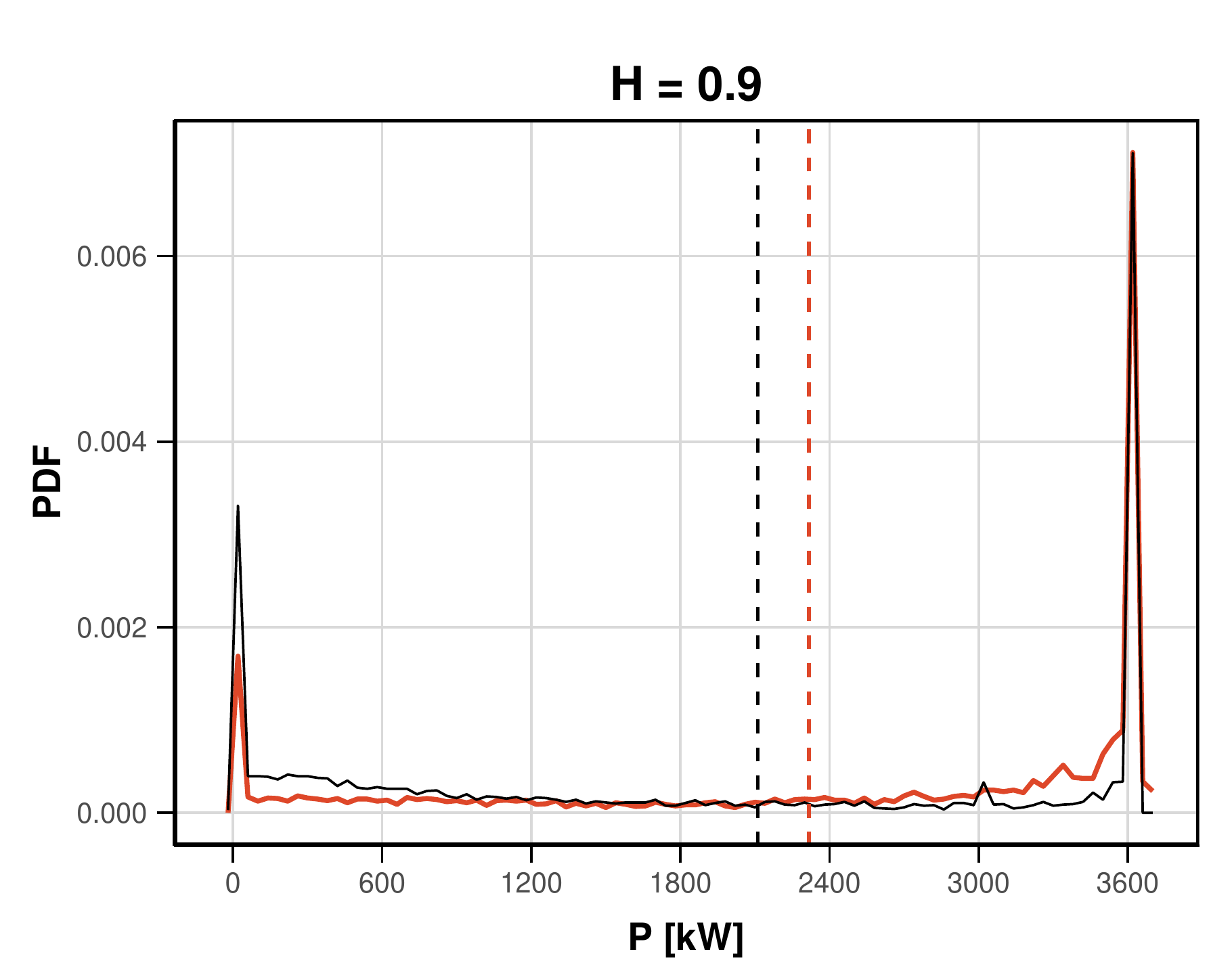}
\caption{Comparison between PDF of empirical (black) and simulated (orange) power time series for one exemplary month. The Hurst exponent $H$ is varied for every histogram. The dashed lines indicate the respective average power values.}
\label{fig6}
\end{figure}

Figure \ref{fig5} shows three randomly chosen numerical time series $P(t)$ with the different Hurst exponents and compares them to one empirical time series at the bottom. A visual inspection gives a first indication of 
the similarity between the empirical results from sec. \ref{sec3} and the modelled time series.
For $H=0.9$ the modelled time series reproduces both the transitions between $P_-$ and $P_+$ and the fluctuations around local trends most accurately. The strong abrupt ramps on short time scales can also be observed for all modelled time series. Note that if power time series had uncorrelated fluctuations (red curve), they would apparently tend to change their local trend more frequently. 
After this first visual comparison we next present a more quantitative comparison of model and empirical data. Therefore we consider the bimodality of power statistics, the intermittent behaviour of the increment statistics and both the autocorrelation and power spectra of power time series.
\FloatBarrier
The quality of the reproduced PDF shown in fig.\ref{fig6} for different values of $H$ is not as obvious. This time, we choose the sampled time series $P(t)$ whose PDF reproduces the height of the bimodal peaks most sufficiently but still reflects a typical result. While the PDF for $H=0.9$ reproduces the bimodality of the power time series in a satisfactory manner, the average values (dashed line) clearly differ. For $H=0.5$, peak heights can not be reproduced: a sample of $P(t)$ with uncorrelated fluctuations tends to occupy both fixed points $P_-$ and $P_+$ with the same frequency. This behaviour is entailed by the enhanced frequency of transitions we observed in fig.\ref{fig5}. Yet it fails to give a convincing result for the mean value even though it should be more stably centered for a balanced PDF.

\begin{figure}[!h]
\centering
\includegraphics[width=10.5cm]{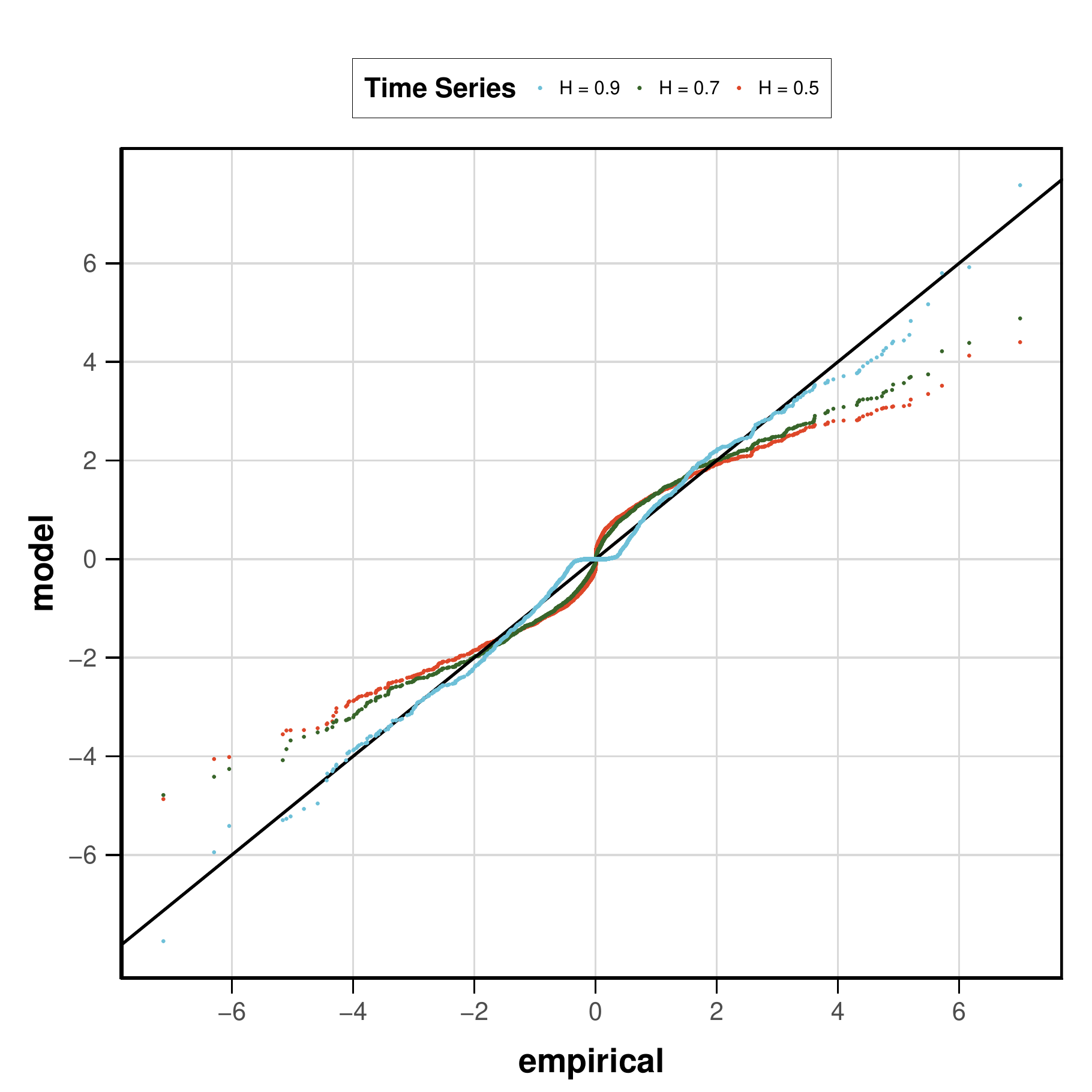}
\caption{Direct comparison between empirical and simulated distributions of power increments $\Xi(t)$ in a QQ--plot for different Hurst exponents $H$. The ordinate shows the quantiles of the modelled distribution, the abscissa those of the empirical distribution. The black diagonal line serves as a reference for an equal distribution of both.}
\label{fig7}
\end{figure}

\begin{figure}[!h]
\centering
\includegraphics[width=10.5cm]{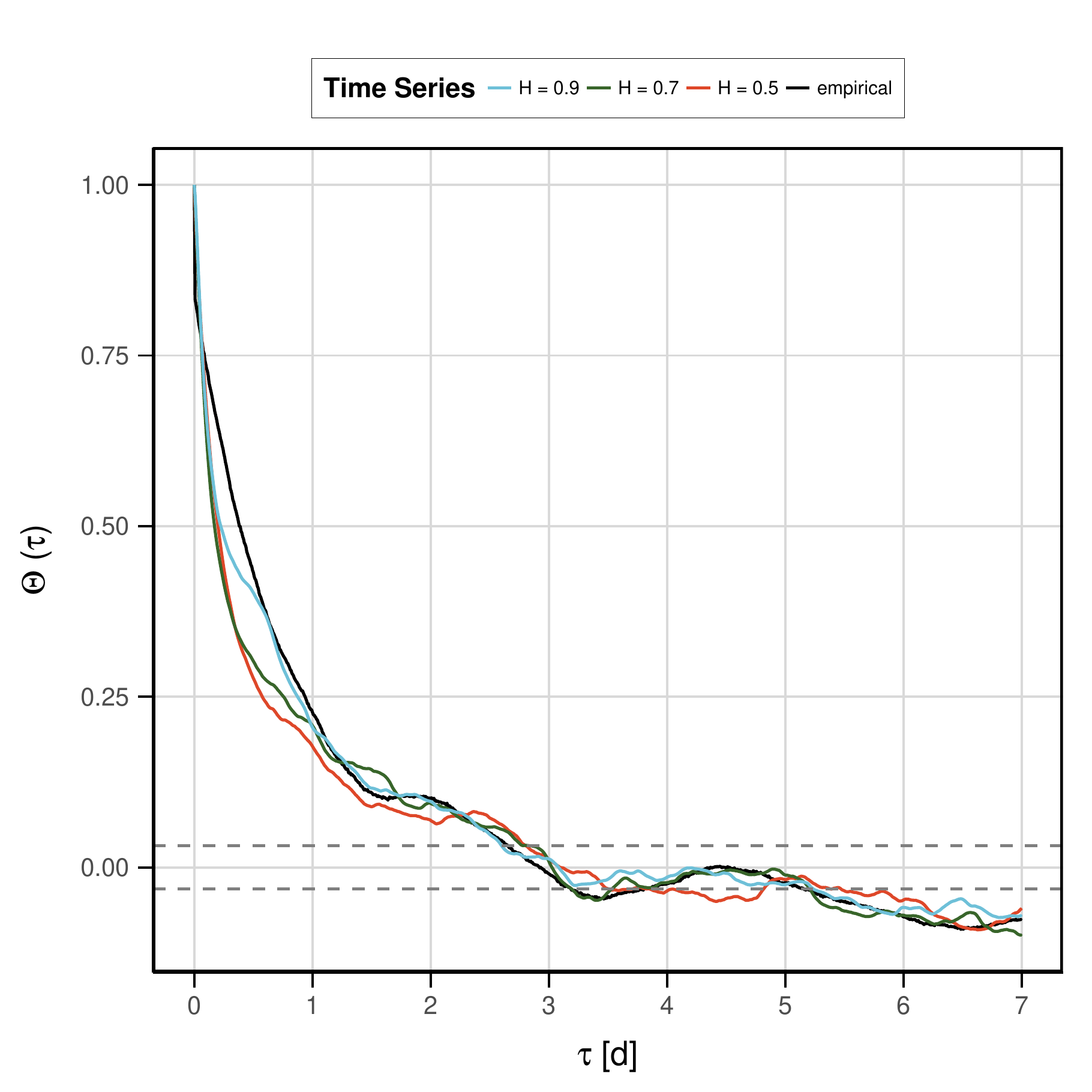}
\caption{Comparison between autocorrelation $\Theta(\tau)$ of empirical and simulated power time series for different Hurst exponents $H$. The maximum delay $\tau$ is one week. The dashed horizontal lines represent the averaged level of significance.}
\label{fig8}
\end{figure}

The power increments $\Xi_k(t)$ are an essential quantity for the control of WTs and grid stability. Besides the analysis in terms of autocorrelation and power spectra of $P_k(t)$ (see below), higher statistical features of the power fluctuations can be grasped by investigating the statistics of the power increments, yielding higher order statistical features of power fluctuations. 
A comparison of increment statistics between real and modelled data can display strengths and weaknesses of the model to capture the intermittent nature of power generation, expressed by strong correlated fluctuations. To this extent, we compare the quantiles of the model results for the power increments with the empirical data. This empirical test clarifies qualitatively in which way the two distributions generally differ regarding their quantiles.
\FloatBarrier
Since our model captures the effect of different degrees of autocorrelation for the power time series with the Hurst parameter $H$, we can also study the effect of autocorrelation in the power time series on the increment statistics and possible non-Gaussian characteristics. In fig.\ref{fig7}, the quantiles of three simulated power time series with varied $H$ are plotted against the quantiles of an empricial power time series. If they were equally distributed, they would follow the diagonal black line. It becomes apparent that in the strongly persistent regime of $H=0.9$, the larger increments which possess the non-Gaussian property are reproduced most sufficiently. Increments up to $|\Xi_k(t)| < 4\sigma$ are almost equally distributed to the randomly chosen empirical time series. The deviations mostly occur in the heavy tails of the distributions for larger absolute increments. Small increments $|\Xi_k(t)| < \sigma$ are not adequately described by any of the sampled time series. We point out that the shape of the QQ--curve for persistent samples $P(t)$ with $H=0.7$ differs only slightly from the one with $H=0.5$. 
\FloatBarrier
From this we conclude that a certain threshold of persistence has to be exceeded so that big jumps between $P_-$ and $P_+$ dominate and generate strong heavy tails. Thus, the persistence of power is closely related to its intermittent behaviour regarding its fluctuations.
The results we compared so far demonstrated that the model succesfully reproduces important statistical characteristics of power generation with respect to the statistical distribution. The variation of Hurst exponents as a tool to incorporate the temporal correlation of power time series should yet yield a strong impact on the autocorrelation function. Thus, we analyse how well the autocorrelation function $\Theta(\tau)$ and also the power spectrum is reproduced by our model.  Figure \ref{fig8} shows in black the monthly averaged autocorrelation of empirical power time series. We compare autocorrelations for the three different Hurst exponents in the same average. The model in general succeeds to reproduce the slowly decreasing autocorrelation with a decay on the same time scale.
Although we observe the expected relation that higher values of $H$ lead to a more slowly decaying autocorrelation, none of the model results yields a sufficient reproduction of the decay for $\tau \le 1\, \mathrm{d}$. Only for $H = 0.9$ the simulated autocorrelation approaches the empirical one within this interval. In general, the results for $\tau > 1\,\mathrm{d}$ are more applicable and even tend to reproduce the weak anticorrelation for $\tau > 5\,\mathrm{d}$. The discussed seasonal bumps are not clearly distinguishable for any of the model results. The model autocorrelations show stronger fluctuations which indicates that the monthly average does not cancel out the impact of monthly varying time scales as much as for the empirical data. A more sophisticated approach for the seasonal component in our model could enhance this aspect. 

\FloatBarrier

Finally, fig.\ref{fig3b} shows the three corresponding averaged power spectra for the model time series (green). In contrast to the empirical spectrum, the slope remains almost constant for all $H$ values even beyond the cut-off value (dashed line), yielding lower power for the higher frequencies. Within the fitted region, all three spectra show a qualitatively similar decay with $\beta \approx 5/3$. In particular, the time series with the highest persistence again gives the most convincing result ($H=0.5: \, \beta = 1.71 \pm 0.03\, , \ H=0.7: \, \beta = 1.69 \pm 0.04\, ,$ $H=0.9: \, \beta = 1.64 \pm 0.03$).

\section{Conclusion}
\label{sec5}
We analysed a data set comprising time series of 30 WTs located at the German offshore windfarm  RIFFGAT over a total time period of one year. As important basic features, we found power values of each WT to be bimodally distributed with density peaks at zero power output $P_-$ and rated power output $P_+$. The power output $P_k(T)$ of a WT turned out to have a slowly decreasing autocorrelation which connects to general findings in the literature. Beyond these basic features, we took into account the non--stationarity of time series by applying DFA to both $P_k(T)$. In terms of the Hurst exponent as a general classifier for autocorrelations of single WTs, different characteristic time scales were identified on which correlations vary between persistence and more involved nonstationarity, separated by a crossover $s_c \approx 3\,\mathrm{d}$.
Based on these findings, we put forward a reduced form model for the power output $P_k(t)$. With this model we intended to provide a tool for reproducing several statistical properties and testing power generation scenarios with respect to empirically motivated parameters. To this extent, we employed a nonlinear stochastic differential equation with a double--well potential and a correlated diffusion component. Most parameters were estimated empirically while some were optimized to capture statistical properties of the data. The model succeeded to qualitatively reproduce important statistical characteristics such as a bimodal PDF, intermittent fluctuations and a slowly decreasing autocorrelation. It thus enriches the modelling perspective on power generation by focussing on a comprehensive set of statistical features only while still producing characteristical results. Possible applications include wind farm or large power network simulations that still aim at reflecting single WT dynamics adequately. Even for such aggregated large-scale approaches, taking correlations on a single WT scale into account is essential since models with uncorrelated fluctuations will underestimate important statistical effects such as heavy tailed distributions and persistence.

Anyway, the proposed model should be regarded as a first step towards a study that quantitatively reflects on the model dynamics in greater detail and undermines the significance of the obtained results. While a systematic examination of parameter effects on the outcome remains to be done, some parameter estimation methods could also be subject to further improvement. The observed crossover which uncovers different correlations on time scales $t\le3\,\mathrm{d}$ could be of interest for possible model extensions. The strong ramping behaviour on short time scales that is suspected to entail the found crossover is an important component in the context of predictability despite strong intermittent fluctuations. Finally, we only focused on single WTs. An evident extension of the proposed model would be to aggregate several correlated single WTs' \cite{chen2010stochastic} and wind farms' \cite{guan2018multiple} model power outputs and study characteristics of aggregated output \cite{chen2010stochastic}, especially with regards to an amplification of correlated fluctuations on the aggregated level.

\section*{Acknowledgments}
Parts of this work have been financially supported by the
Ministry for Science and Culture of the Federal State of Lower
Saxony within the project ’Ventus Efficiens’ under grant num-
ber ZN3024. We further acknowledge support by the Open Access Publication Fund of the University of Duisburg-Essen.

\section*{Conflict of interest}
The authors declare that they have no conflict of interest.

\section*{Appendix: Parameter Estimation}
\label{app}
We briefly provide an overview of the model parameters and explain the methods of how they are calibrated. To solve the SDE, we use the Euler-–Maruyama--method for SDEs with colored noise \cite{generatefgn}. We sample the latter by using the so called Harte--method, applying the Durbin--Levinson--algorithm \cite{durbinlevinson}. Averaged over 10000 samples, it takes us $0.095\pm 0.009\,\mathrm{s}$ to simulate a power time series covering a period of one month with the hardware setup used in this study. 
One advantage of our model lies in the fact that we can already fix six of ten parameters before we simulate a single time series only by adjusting them in a particular way to the empirical data. Consequently, the estimation of these six fitted parameters is enhanced the more data is included. This data based approach enables us to assign interpretations to the parameters in context of the empirical data.

\begin{itemize}
\item Center of potential function $P_0$:\\
The most apparent value for the center of the symmetric double--well potential $P_0$ is to fix it at the center of the empirical range of power values $P_0 = 1800\,\mathrm{kW}$. This choice simply ensures that for an approriate choice of $a$, the model reproduces the bimodal peaks at the correct power output.

\item Seasonal frequency $\omega$:\\
Since we introduced the frequency $\omega$ of the periodical driving force into our model to include seasonal variations, we estimate it from the empirical wind speed time series $u(t)$ by applying Fourier analysis. By choosing $\omega$ to be the Fourier frequency of the highest Fourier amplitude, we make sure that it at least resembles the most dominant seasonal frequency in the data. We choose the wind speed time series $u(t)$ instead of the power times $P(t)$ because we regard $u(t)$ to be the external drive for the WTs analogously to an external driving force of a periodically driven oscillator. This simple approach can only account for a rough estimation of the seasonal variations.

\item Probabilities for excess/negative power $p_{\gtrless}$:\\
Both parameters need to be set with respect to the likelihood of empirical power values beyond the operating range. A power lower than $P_-$ or higher than $P_+$ can only be generated if the corresponding wind speed $u$ is lower than $u_-$ or higher than $u_+$ respectively. Hence we calculate the probabilities for such wind speed values from the empirical data and use these probabilities to fix the model probabilities $p_{\gtrless}$. Since these probabilities show significant variations between different months, we obtain 24 values in total that can be interpreted as the monthly likelihood for the WT to generate power beyond its operating range.

\item Standard deviation for excess/negative power $\sigma_{\gtrless}$:\\
How strongly does power fluctuate around $P_-$ and $P_+$ when this range is exceeded? --- As we already stated, we choose a simple Gaussian distribution for the power values around these two values. We fix $\sigma_{\gtrless}$ applying a Gaussian empirical fit around $P_-$ and $P_+$ respectively. Since this empirical standard deviation hardly varies for different month, we only obtain the two values $\sigma_+ = 68.93\,\mathrm{kW}$ and $\sigma_- = 4.47\,\mathrm{kW}$.
\end{itemize}
The remaining four parameters should be regarded as more dynamical parameters that we try to estimate by optimizing a certain statistical feature with respect to this parameter. One exception is the Hurst exponent $H$ which we vary between the three values $H = 0.5 \, , \, 0.7 \, , \, 0.9$ in sec. \ref{sec4.2} towards a better understanding of the impact of power law autocorrelations on power generation. We briefly sketch the estimation methods for the parameter $a$ of the potential function, the diffusion strength $D$ and the strength of the seasonal variations $A$. For each parameter, we optimize it by sampling a reasonable number of time series for each of the 12 months.
\begin{itemize}
\item Curtailment indicator $a$:\\
After we have set $P_0 = 1800\,\mathrm{kW}$, the most obvious choice for $a$ would also be $a = 1800\,\mathrm{kW}$ to ensure that the bimodal peaks center around the correct fixed points $P_- = 0\,\mathrm{kW}$ and $P_+ = 3600\,\mathrm{kW}$. Anyway, the artificial flattening procedure plays an important role in this context: if we choose $a = 1800\,\mathrm{kW}$, the amount of values that are either set to one of the thresholds $P_-$ or $P_+$ or scattered around them is too high compared to the empirical data. This hints at choosing a value $a < 1800\,\mathrm{kW}$. We optimize $a$ by sampling power time series with different $a < 1800\,\mathrm{kW}$ and calculating the amount $p_{\gtrless}$ of power values above (below) or around $P_+$ ($P_-$). The resulting value (for fixed $H$) minimizes the difference between the respective empirical and simulated value. Consequently, we regard the difference $\,(P_0 - a)\,$ as the power difference that entails how strongly the empirical power time series are subject to curtailment.

\item Diffusion strength $D$:\\
The constant diffusion strength $D$ determines how strongly the stochastic component dominates power generation. It is an important parameter in the context of transition time scales between the two fixed points $P_-$ and $P_+$. If we regard $D\xi_H(t)$ as the diffusion function of a Langevin equation, we can apply the standard method to obtain this diffusion function directly from the empirical data \cite{empiricalDD}. It can be shown \cite{langevin} that the diffusion function can be extracted by calculating

\begin{align}
D^{(2)}(P,t) \ &= \ \lim\limits_{\tau \rightarrow 0}{\frac{\langle(P(t+\tau) - P(t))^2\rangle}{\tau}\bigg\rvert_{P(t)=P_0}} \, .
\end{align}

Since this method potentially yields errorenous results for correlated noise processes, we can only expect an approximate estimate of $D$ \cite{ddheart}. Nevertheless, we follow this approach by minimizing the difference $\mathrm{min}_D\left[ |D_2^{\mathrm{emp}}(P(t)) - D_2^{\mathrm{sim}}(P(t);D)|\right]$ of the empirical and the model diffusion function with respect to $D$.

\item Seasonal variation strength $A$:\\
$A$ biases the power generation towards one of the two peaks of the bimodal power distribution. In winter months, a high average power output is likely whereas in spring, usually lower wind speeds occur which lead to a pronounced peak around $P_-$. Thus we can calculate
 
\begin{align}
E(t) = \sum\limits_{t'=t_0}^t P(t')
\end{align}

resembling an energy indicator for the empirical and simulated data to minimize the difference between the respective time-averaged values $\langle E(t)\rangle_t$. We consider $A$ to be optimized when this difference of averaged accumulated power values is minimized for each month.
\end{itemize}
We summarize the parameters choices we obtain from the above mentioned estimation methods:

\begin{table}[h]
\begin{center}
\caption[]{Estimated parameters for different Hurst exponents $H$.}
\begin{tabular}{|c||c|c|c|}
\hline 
$\mathbf{H}$ & $\mathbf{a \ [\mathrm{kW}]}$ & $\mathbf{D \ [\mathrm{kW} / \sqrt{\mathrm{s}}]}$ & $\mathbf{A \ [\mathrm{kW} / \mathrm{s}]}$  \\ 
\hline \hline
\textbf{0.5} & 1755 & 355 & 0.15  \\ 
\hline 
\textbf{0.7} & 1445 & 360 & 0.17  \\ 
\hline 
\textbf{0.9} & 1235 & 485 & 0.26  \\ 
\hline 
\end{tabular} 
\label{tab1}
\end{center}
\end{table}
\FloatBarrier

\bibliographystyle{ieeetr}
\bibliography{bibli}  %%% Uncomment this line and comment out the ``thebibliography'' section below to use the external .bib file (using bibtex) .

%%% Uncomment this section and comment out the \bibliography{references} line above to use inline references.
% \begin{thebibliography}{1}

% 	\bibitem{kour2014real}
% 	George Kour and Raid Saabne.
% 	\newblock Real-time segmentation of on-line handwritten arabic script.
% 	\newblock In {\em Frontiers in Handwriting Recognition (ICFHR), 2014 14th
% 			International Conference on}, pages 417--422. IEEE, 2014.

% 	\bibitem{kour2014fast}
% 	George Kour and Raid Saabne.
% 	\newblock Fast classification of handwritten on-line arabic characters.
% 	\newblock In {\em Soft Computing and Pattern Recognition (SoCPaR), 2014 6th
% 			International Conference of}, pages 312--318. IEEE, 2014.

% 	\bibitem{hadash2018estimate}
% 	Guy Hadash, Einat Kermany, Boaz Carmeli, Ofer Lavi, George Kour, and Alon
% 	Jacovi.
% 	\newblock Estimate and replace: A novel approach to integrating deep neural
% 	networks with existing applications.
% 	\newblock {\em arXiv preprint arXiv:1804.09028}, 2018.

% \end{thebibliography}

\end{document}